\documentclass[letterpaper,twocolumn,10pt]{article}
\usepackage{usenix-2020-09}

\usepackage{tikz}
\usepackage{amsmath}
\usepackage{filecontents}
\usepackage{times}  
\usepackage{graphicx}
\usepackage{xspace}
\usepackage{xcolor}
\usepackage[T1]{fontenc}
\usepackage{caption}
\usepackage{subcaption}
\usepackage{verbatim}
\usepackage{url}
\usepackage{enumitem}
\usepackage{listings}
\usepackage{multirow}
\usepackage[ruled,linesnumbered]{algorithm2e}
\usepackage{pifont}
\usepackage{mathtools}
\usepackage{fancyhdr}
\usepackage{xurl}
\usepackage{textcomp}
\usepackage{pifont} % for \ding{}
\usepackage{afterpage}
\usepackage{cite}
\usepackage[export]{adjustbox}

\usepackage[capitalise,noabbrev]{cleveref} % clever reference
\crefformat{section}{\S#2#1#3}

\newcommand{\sysname}{CascadeInfer\xspace} % short for Low-Latency, Load-Balanced, Length-Adaptive, LLM Inference

\newif\ifprintcomments           
\printcommentstrue

\newcommand{\parab}[1]{\noindent\textbf{#1}}

\newcommand{\blackcircle}[1]{%
\ifcase#1\relax
\or \ding{182}% 1
\or \ding{183}% 2
\or \ding{184}% 3
\or \ding{185}% 4
\or \ding{186}% 5
\or \ding{187}% 6
\or \ding{188}% 7
\or \ding{189}% 8
\or \ding{190}% 9
\or \ding{191}%10
\else ??\fi% 
}

\usepackage[justification=RaggedRight,labelfont=bf,font=small,textfont=normalfont,tableposition=below]{caption}

\begin{document}

\date{}

% \title{\sysname: \underline Low-Latency and \underline Load-Balanced \underline  LLM Serving\\ via \underline Length-Aware Scheduling}

\title{\sysname: Length-Aware Scheduling of LLM Serving\\with Low Latency and Load Balancing}

\author{
{\rm Yitao Yuan\textsuperscript{1,2}\thanks{This work was done during an internship at ScitiX AI.}},
{\rm Chenqi Zhao\textsuperscript{1}},
{\rm Bohan Zhao\textsuperscript{2}},
{\rm Zane Cao\textsuperscript{2}},
\\
{\rm Yongchao He\textsuperscript{2}\thanks{Corresponding authors.}},
{\rm Wenfei Wu\textsuperscript{1}\hyperlink{Hfootnote.2}{\textsuperscript{\textdagger}}}
\vspace{0.2cm} \\
% \\
{\rm \textsuperscript{1}Peking University} \ \ \ \ \ 
{\rm \textsuperscript{2}ScitiX AI}
}

\maketitle

\begin{abstract}
Efficiently harnessing GPU compute is critical to improving user experience and reducing operational costs in large language model (LLM) services. However, current inference engine schedulers overlook the attention backend's sensitivity to request-length heterogeneity within a batch. As state-of-the-art models now support context windows exceeding 128K tokens, this once-tolerable inefficiency has escalated into a primary system bottleneck, causing severe performance degradation through GPU underutilization and increased latency. We present \sysname, a runtime system that dynamically reschedules requests across multiple instances serving the same LLM to mitigate per-instance length heterogeneity. \sysname partitions these instances into length-specialized groups, each handling requests within a designated length range, naturally forming a pipeline as requests flow through them. \sysname devises a dynamic programming algorithm to efficiently find the stage partition with the best QoE, employs runtime range refinement together with decentralized load (re)balance both across and within groups, achieving a balanced and efficient multi-instance service. Our evaluation shows that, under the same configuration, \sysname reduces end-to-end latency by up to 67\% and tail latency by up to 69\%, while improving overall system throughput by up to 2.89$\times$ compared to the state-of-the-art multi-instance scheduling systems. 
\end{abstract}

\section{Introduction}
\label{sec:introduction}

Large Language Models (LLMs) are increasingly deployed in production to support long-context, multi-turn applications such as chat assistants and agents~\cite{chatgpt, grok}. To sustain massive request volumes, providers operate clusters of replicated inference instances (e.g., vLLM~\cite{kwon2023vllm}, SGLang~\cite{sglang}), forming \emph{multi-instance LLM serving} (\textbf{MILS}) systems. 

\noindent
\textbf{A growing mismatch.}
In MILS deployments, each instance executes requests in fine-grained decoding iterations using continuous batching~\cite{yu2022orca}. Under short-context workloads, this abstraction worked well: batching available requests was sufficient to achieve high GPU utilization.  
However, modern long-context models routinely support token windows of 128K or more~\cite{llama-3.2, liu2024deepseekv3, chatgpt}. With long-context workloads, we observe a fundamental mismatch between this \emph{scheduling abstraction} and the \emph{hardware characteristics} of attention kernels. Our analysis of production traces and public dialogue datasets~\cite{shareGPT} shows highly skewed request-length distributions: many short requests coexist with a small but increasingly common set of extremely long ones.

Attention computation—whose cost scales with sequence length—dominates decoding latency not only for a single long request (up to 80\% for length at 64K context length~\cite{yuan2025native}) but also for large batches (up to 81\% for 1K sequences with a batch size of 250, as demonstrated in \cref{ssec:serving_model}).
 Critically, modern attention backends exhibit strong performance sensitivity to \emph{length heterogeneity}: when a batch mixes short and long sequences, GPU utilization collapses due to unavoidable load imbalance and synchronization stalls. These inefficiencies stem from fundamental hardware execution constraints and persist even with advanced kernels such as FlashAttention~\cite{dao2022flashattention} and Triton-based implementations~\cite{tillet2019triton,sglang}. In short, \emph{the attention backend wants homogeneous lengths, but the serving stack keeps giving it heterogeneous batches}.

\noindent
\textbf{Why existing schedulers fall short.}
State-of-the-art (SOTA) serving systems attempt to hide heterogeneity through intra-instance scheduling~\cite{kwon2023vllm, sglang, wu2023fast, cao2025locality}. Yet an instance's memory capacity and latency constraints fundamentally limit its ability to buffer enough requests to form homogeneous batches—an inherent consequence of Little's Law~\cite{little1961proof}.  
Inter-instance schedulers (e.g., Llumnix~\cite{sun2024llumnix}) distribute requests across instances to balance load or memory usage, but they are \emph{length-agnostic}. As a result, every instance still receives heterogeneous batches and suffers the same backend inefficiency. No existing solution reshapes global request flow to match the backend's computational needs.

We argue that the remedy must occur above the instance level. We observe that the massive scale of MILS, while typically viewed as a source of complexity, actually offers the solution to this problem. 
By scheduling requests among global instances, \sysname turns the scale of MILS from a management burden into an optimization opportunity. Guided by this insight and the strong backend preference for homogeneous lengths, we propose to restructure MILS scheduling around the principle of: \emph{``Global homogenization for local efficiency.''}
Instead of having every instance cope with global heterogeneity, we reshape the cluster into a \emph{length-aware pipeline} where each instance (or group of instances) processes requests of similar length. As requests grow during decoding, they move downstream through the pipeline, so every instance naturally forms length-homogeneous batches \emph{without modifying} its internal scheduler.

This paper presents \sysname, a decentralized, length-aware inter-instance scheduling system for MILS clusters.  
\sysname plans and maintains a multi-stage pipeline over the sequence-length space, routes incoming requests to the appropriate stage, and migrates them as they grow. To adapt to workload dynamics, \sysname performs:
\emph{(1) Workload-aware stage boundary adjustment}: neighboring instances exchange lightweight length summaries and recompute boundaries to minimize predicted latency.
\emph{(2) Decentralized bid-ask scheduling}: overloaded instance offer requests while underloaded peers bid to acquire them, equalizing intra-stage load without a central coordinator.

Across a 16-GPU deployment with production workloads, \sysname reduces median latency by up to 67\%, tail latency by up to 69\%, and increases throughput by up to 2.89$\times$ over SOTA inter-instance schedulers.

In summary, our contributions are as follows:

(1) We identify a fundamental mismatch between existing scheduling abstractions and the attention backend’s sensitivity to length heterogeneity in long-context MILS workloads.

(2) We propose a shift from local-only to global-cooperative scheduling. By enforcing global length-homogeneity, \sysname allows unmodified local schedulers to operate near hardware limits.

(3) We implement \sysname on vLLM and demonstrate substantial latency and throughput improvements over SOTA serving systems.

\section{Background and Motivation}
\label{sec:background}

Modern LLM serving clusters execute requests in a multi-instance setting,
where each instance performs iterative token-by-token decoding using continuous
batching~\cite{yu2022orca, kwon2023vllm, sglang}. 
While this abstraction worked well for short-context workloads, emerging 
long-context applications introduce a fundamental tension between 
\emph{how requests arrive} and \emph{how attention kernels execute}. 
This section summarizes the serving model, highlights the 
kernel-level characteristics that make sequence-length heterogeneity harmful, 
and explains why existing schedulers cannot address the resulting inefficiency. 
We conclude with the design insight that motivates \sysname.

\subsection{Workflow of Inference}

In LLM inference, user requests are executed under \emph{continuous batching}~\cite{kwon2023vllm}. Transformer-based LLMs handle these
requests in two distinct stages: \emph{prefill} and \emph{decoding}.

\noindent\textbf{Prefill.}
The model ingests the entire input sequence of length $n$ in parallel to construct the initial
\emph{key-value (KV) cache}. Prefill involves a full self-attention over the prefix and therefore exhibits
\emph{quadratic} time complexity, $O(n^2)$~\cite{vaswani2017attention}. In practice, serving systems isolate prefill into dedicated iterations to fully utilize compute-bound kernels~\cite{zhong2024distserve}.

\noindent\textbf{Decoding.}
After prefill, the model generates output tokens autoregressively. Modern serving systems batch
multiple active requests into each decoding iteration. Each step reads all previously cached
keys and values, performing an attention operation with \emph{linear} time complexity, $O(n)$,
per request. Critically, for long-context models (e.g., 128K tokens), this attention operation
is \emph{memory-bound} and can dominate iteration latency (up to 81\%). Moreover, attention
kernels are highly sensitive to sequence-length heterogeneity within a batch, suffering from
severe load imbalance and inefficient tiling when processing mixed-length sequences.

Whereas prefill consists of uniform, fully parallel computation, decoding performs iterative
batching over requests with diverse sequence lengths. This heterogeneity disrupts the hardware
efficiency of attention backends, creating a fundamental mismatch between traditional
scheduling policies and the execution characteristics of long-context LLMs.

\subsection{LLM Serving Model in MILS Clusters}
\label{ssec:serving_model}

A MILS deployment consists of $M$ identical 
inference instances operating independently. Each instance repeatedly performs
\textit{iteration-level decoding}: at each decoding step, it forms a batch from 
all active requests, executes a forward pass to produce the next tokens, and 
updates KV caches. Continuous batching~\cite{kwon2023vllm} merges 
newly arrived requests into ongoing iterations to maximize GPU utilization.

\begin{figure}[t]
    \centering
    \includegraphics[width=\linewidth]{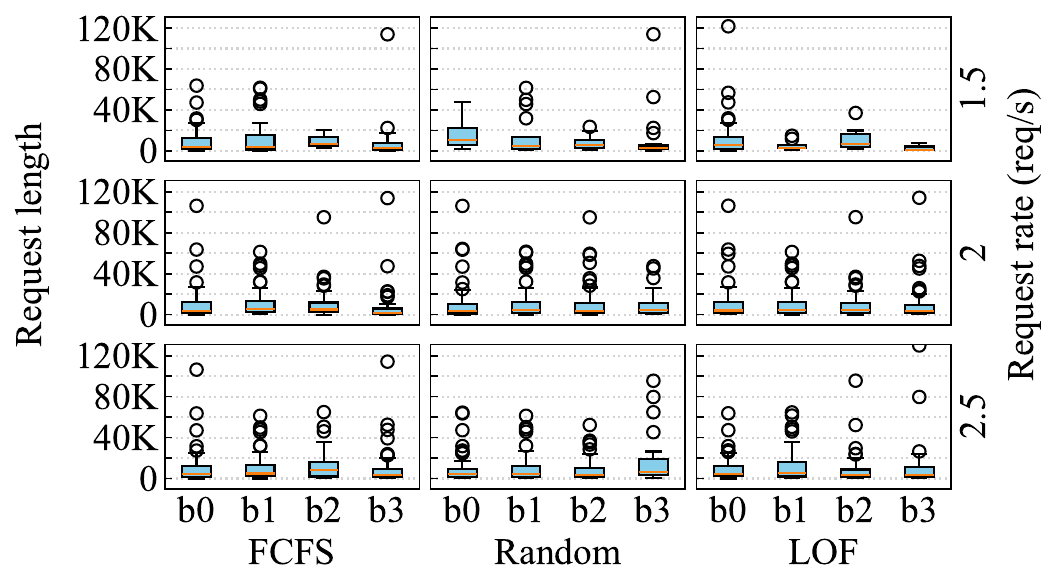}
    \caption{Request-length distribution in batches under various scheduling policies and request rates. Batches were sampled at 20\%, 40\%, 60\%, and 80\% of the inference process. The inputs come from an LLM dialogue dataset~\cite{shareGPT}, and requests longer than 128K are discarded.}
    \label{fig:mixed-len}
\end{figure}

Two properties of this workflow are central to our study:

\emph{(1) Iteration-level latency is dominated by attention.}
          As sequence length and batch size grow, the KV cache becomes comparable to---or even larger than---the model's weight size. Because decoding is memory‐bound, attention overhead then gradually comes to dominate~\cite{yuan2025native, dao2022flashattention}. 
          
          Empirically, our experiments using Llama-3.2-3B on an H100 GPU demonstrate that attention computation dominates latency at large batch sizes, even for short requests. Specifically, for 1000-token sequences, the attention phase accounts for up to 81\% of total latency at a batch size of 250, a sharp increase from 14\% for a single request. Similarly, for 200-token sequences at a batch size of 500, this proportion reaches 62\% (vs.\ 14\% baseline).

\emph{(2) Engines observe highly heterogeneous sequence lengths.}
          Real workloads exhibit skewed length distributions, with many short 
          requests mixed with few but increasingly common long requests 
          (Figure~\ref{fig:mixed-len}). These heterogeneous requests 
          inevitably co-exist in the same decoding iteration.

These properties imply that the batch formed in each iteration---and its 
sequence-length composition---directly determines GPU efficiency.

\textbf{\begin{figure}[t]
    \centering
    \includegraphics[width=0.85\linewidth]{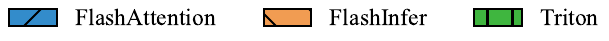}
    \\
    \begin{subfigure}[t]{0.47\linewidth}
        \includegraphics[width=\linewidth]{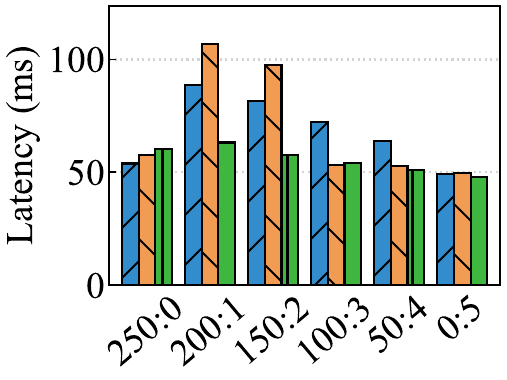}
        \caption{Request length 1000 vs 50000.}
        \label{fig:interference-1}
    \end{subfigure}
    \hfill
    \begin{subfigure}[t]{0.47\linewidth}
        \includegraphics[width=\linewidth]{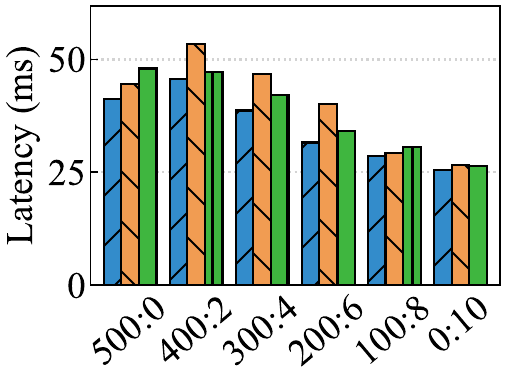}
        \caption{Request length 200 vs 10000.}
        \label{fig:interference-2}
    \end{subfigure}
    \caption{Effect of sequence length heterogeneity on  \emph{decoding forward} pass performance. Measured on a single H100 GPU using vLLM and SGLang with FlashAttention, FlashInfer, and Triton (model: Llama-3.2-3B, batch size: 512).}
    \label{fig:interference}
\end{figure}}

\subsection{Kernel Characteristics: Sensitivity to Length Heterogeneity}
\label{ssec:kernel_characteristics}

Modern attention backends (e.g., FlashAttention, FlashDecoding, Triton-based
implementations) rely on tiling and parallelizing each sequence into 
\textit{blocks} to saturate GPU SMs. The performance of these kernels depends
crucially on the number and shapes of blocks derived from each request’s
sequence length. For a request with length $L$, the kernel generates roughly
$L / B$ blocks (for block size $B$) or uses variable block sizes to fit a
fixed number of blocks~\cite{dao2022flashattention, flashdecoding}.

When a batch mixes short and long sequences, the GPU encounters two sources
of inefficiency:

\emph{(1) Inter-SM imbalance.} Long sequences produce dramatically more blocks than short ones. Even with optimized block partitioning, aggregation and synchronization remain serialized on the critical path of the longest request, leaving SMs waiting after short requests finish.

\emph{(2) Partitioning inefficiency.}
Block-size vs.\ block-count trade-offs~\cite{flashdecoding} cause either (i) excessive aggregation overhead for long sequences (fixed block size), or (ii) poor SM occupancy (oversized blocks under fixed block count). Heterogeneous batches induce both problems simultaneously.

Empirically, batch-level heterogeneity can increase kernel latency to 
1.1--2.1$\times$ even when the total number of input tokens remains constant 
(Figure~\ref{fig:interference}). Crucially, this inefficiency represents
a \emph{hardware-execution property}, not a limitation of a particular serving
implementation. Attention kernels---and thus the GPU---strongly prefer 
\textbf{length-homogeneous} batches. However, MILS workloads naturally produce a heterogeneous mixture, revealing a fundamental mismatch.

\subsection{Why Existing Schedulers Cannot Fix Heterogeneity}
\label{ssec:existing_limits}

Existing schedulers attempt to increase utilization through 
request reordering or routing, yet none can eliminate heterogeneity at the
iteration level.

\noindent \textbf{Intra-instance scheduling.}
Techniques within a single instance~\cite{kwon2023vllm, sglang, fu2024efficient,
cao2025locality} attempt to buffer requests and reorder them for better
batch shape. However, an instance cannot accumulate enough requests to form 
length-homogeneous batches without violating latency targets and exhausting memory. This limitation follows directly from Little’s Law:
$P = \lambda W$. Reducing heterogeneity requires a large queueing population 
($P$), which implies higher memory consumption and longer waiting time
($W$); both are unacceptable.

\noindent \textbf{Inter-instance scheduling.}
Load balancers such as Llumnix~\cite{sun2024llumnix} route requests across
instances based on memory availability or load. These policies are 
length-agnostic: heterogeneity in the global arrival stream is simply 
replicated within every instance. As a result, all instances still process 
mixed-length batches and suffer the same kernel-level inefficiency.

\noindent \textbf{Conclusion.}
The root cause is that current abstractions treat instances as independent 
schedulers. But heterogeneity is a \emph{cluster-level phenomenon} whose 
performance implications arise at the \emph{kernel execution level}. Local 
scheduling cannot fundamentally reshape the global flow of requests to match 
the hardware’s preference.

\subsection{Insight and Challenges}
\label{ssec:insight}

Our key insight is that MILS clusters should be structured into a 
\textbf{length-aware pipeline}: partitioning instances into stages that each 
process a contiguous segment of the sequence-length space. Requests naturally 
grow over time and can be migrated between stages, allowing each instance to 
observe mostly length-homogeneous batches without changing its internal 
scheduler. 
As this approach remains agnostic to intra-instance scheduling strategies, it ensures long-term compatibility with the instance internals.

Designing such a system, however, presents three challenges:

\textbf{Challenge 1: The search space for optimal stage partitioning is explosive.}
Range boundaries determine both performance and balance. The search space is large, which includes the number of stages, the number of instances in each stage, and the sequence length range for each stage; it also depends on workload distributions and kernel behavior. Exhausting all possible combinations cannot find the optimal stage partition within an acceptable time. We define a quality model to quantify the effectiveness of running a batch across multiple instances, and devise a dynamic programming algorithm to search for the partition (batch-to-instances assignment) with the optimal quality in polynomial time (\S\ref{ssec:pipeline_planning}).

\textbf{Challenge 2: The sequence length distribution can drift at runtime and become misaligned with the stage partition, leading to suboptimal performance.}
Stage partitioning is performed based on long-term workload statistics. But workloads and sequence-length distributions evolve over time. The misalignment between the partition and the distribution leads to performance degradation, but frequently (re)performing stage partitioning incurs significant overhead. 
Stage ranges must adapt to the distribution without disrupting ongoing decoding.
\sysname collects runtime length distribution and refines stage ranges to optimize the service quality (\S\ref{ssec:boundary_adjustment}).

\textbf{Challenge 3: Loads of instances in the same stage could evolve to be imbalanced, and the system needs a scalable way to rebalance them.}
Intra-stage load imbalance stems from inherent request diversity, compounded by stale state information and the irregularity of inter-stage request migrations. Na\"ively applying a central scheduler to balance and migrate requests introduces latency and scalability issues, canceling out its performance gains.
\sysname devises a decentralized bid-ask protocol: instances negotiate request migrations via sender-receiver bids, enabling efficient, scalable workload rebalancing (\S\ref{ssec:bid_ask}).

\section{\sysname Overview}
\label{sec:overview}

\subsection{Architecture}
\label{sec:archi}

When serving an LLM, \sysname organizes inference instances into multiple \emph{length-specialized groups}, each dedicated to sequences within a specific length range. During autoregressive decoding, sequence lengths increase monotonically. \sysname leverages this property to form a logical \emph{pipeline}, with each group as a pipeline stage; as decoding progresses, sequences naturally flow from shorter to longer stages. As shown in \cref{fig:arch}, this pipeline comprises $N$ stages, each with a variable number of instances. Within each instance, \sysname retains standard components: a \emph{scheduler} for sequence batching and a \emph{GPU executor} for batch processing on the attention backend. To support adaptive, decentralized rescheduling, \sysname introduces two lightweight runtime components—\textit{LoadTracker} and \textit{Coordinator}—in each instance.

The LoadTracker monitors the instance's token-level workload (e.g., the number of input and output tokens per request) and exchanges these statistics with other instances—both peers in the same stage and instances in adjacent stages. This real-time monitoring allows \sysname to capture transient workload fluctuations and trigger responsive adjustments.
The Coordinator manages dynamic sequence migration by transferring metadata and KV cache across instances, either within a stage or to the next stage. By operating in a decentralized manner, the Coordinator eliminates the bottleneck of a central scheduler, thereby ensuring cluster scalability and minimizing the risk of scheduling decisions based on stale state information.

\subsection{Workflow}

\parab{System bootup.}
\sysname constructs the instance pipeline at startup. 
It applies a length-aware stage partition algorithm (\S\ref{ssec:pipeline_planning}) to partition instances into length-specialized stages. Each instance initializes its runtime components and establishes connections to peer instances within the same stage and to instances in adjacent stages. These connections enable both intra-stage and cross-stage coordination.

\begin{figure}[t]
    \centering
    \includegraphics[width=\linewidth]{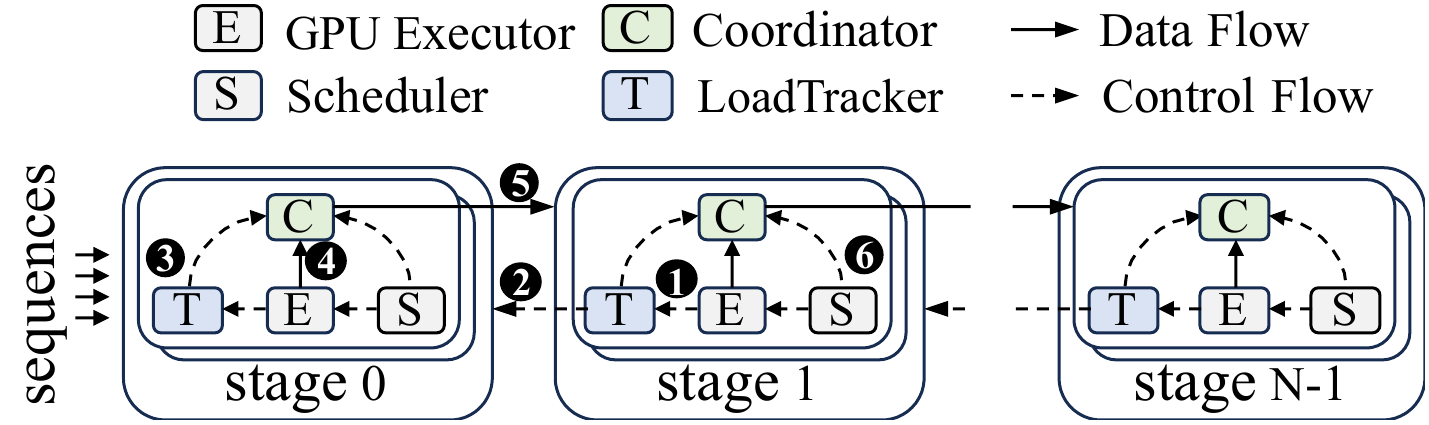}
    \caption{Architecture and workflow of \sysname. Engine instances are grouped by length into stages forming a logical pipeline; sequences may exit early without traversing all stages.}
    \label{fig:arch}
\end{figure}

\parab{Runtime request processing.} \sysname performs progressive decoding with stage transitions as described in \cref{ssec:serving_model}.
As shown in \cref{fig:arch}, when a request arrives, it is routed to the earliest stage whose serving range covers its initial length. The assigned instance adds this request to its batch when memory is available, then performs prefill and decoding locally using its scheduler and GPU executor. As the sequence grows, if its length exceeds the instance's range, it is migrated to the next stage for further decoding. This process repeats until the sequence completes.

\parab{Runtime load (re)balancing.} Each LoadTracker monitors token-level GPU usage (\ding{182}) and propagates this information to peer instances in the same stage and to the upstream stage (\ding{183}). The LoadTracker triggers periodic range refinement and instantaneous request migration.

With the load record over a period, an instance's LoadTracker initiates stage range refinement. It collects the requests of itself and its successor instances. By running the range-refinement algorithm in \cref{ssec:boundary_adjustment}, it obtains a new boundary suitable for the current length distribution.

With instantaneous instance loads (of peers and successors), when an instance hands requests to its successors, or when it finds itself overloaded within the same stage, the instance's LoadTracker triggers the Coordinator (\ding{184}) to (re)balance load by migrating active requests within or across stages. The Coordinator runs a bid-ask protocol to determine the target instance for each request.
For each migration, the Coordinator consults the scheduler to extract sequence metadata (e.g., token count, KV cache location and size, \ding{187}), allocates memory on the target instance, and transfers the KV cache and control state asynchronously (\ding{185}\ding{186}) via \emph{live migration}~\cite{sun2024llumnix, fu2024serverlessllm} (\cref{ssec:bid_ask}).
\section{\sysname Mechanisms}
\label{sec:design}
We first introduce a service quality model to represent the performance of an engine instance on a given batch (\cref{ssec:model}), then construct an initial MILS pipeline using historical workload statistics (\cref{ssec:pipeline_planning}). We devise a runtime pipeline stage boundary adjustment to balance inter-stage load (\cref{ssec:boundary_adjustment}) and decentralized request (re)scheduling to balance intra-stage load (\cref{ssec:bid_ask}).

\subsection{QoE Model for Requests and Instances}
\label{ssec:model}

\parab{QoE model for a request.} After submitting a request, the user endures a prefill delay for the first token, then receives one token per iteration. We define the user's quality of experience (QoE) as a combination of these two times:
$
Q = \alpha  T^{\text{Prefill}} + \beta  T^{\text{Iteration}},
$
where $T^{\text{Prefill}}$ and $T^{\text{Iteration}}$ are the two core serving-quality metrics---Time to First Token (TTFT) and Time per Output Token (TPOT), respectively.
Since the prefill phase costs $O(I^2)$ time for an input request of length $I$, and each decoding iteration costs $O(L)$ time for a request of length $L$, 
we can express $Q$ as:
$$
Q = \left(C_0 + C_1I + C_2I^2\right) + \left(C_3+C_4L\right),
$$
where the first term denotes the quadratic time complexity of prefill and the second term denotes the linear time complexity of one iteration. $C_i$ are coefficients fitted by experiments.

\parab{QoE model for a batch.} When an instance processes a batch $B$ (with $n$ requests), each request's processing time is stretched to the same value, namely for any $ j\in \{1,\dots,n\} $:
{\small
$$
Q_j = \left(C'_0 + C''_0 n + C_1\sum_{i=1}^n I_i + C_2\sum_{i=1}^n I_i^2\right) + \left(C'_3+C''_3 n + C_4\sum_{i=1}^n L_i\right),
$$
}
$$
(C'_0+C''_0=C_0, C'_3+C''_3=C_3) ,
$$
where variables with subscript $i$ denote the attributes for request $i$. The overall batch QoE is defined as 
\begin{equation}
  \label{eq:batch_quality}
Q^B = \sum_{j=1}^n Q_j = nQ_1 = n\sum_{k=0}^4 D_kF_k,
\end{equation}
where $D_k$ are coefficients, $F_0=1$, $F_1=n$, $F_2=\sum_{i=1}^nI_i$, $F_3=\sum_{i=1}^nI_i^2$, and $F_4=\sum_{i=1}^nL_i$.

\parab{Fitting the parameters $D_k$.} We run experiments with various batch settings. For each run, we obtain $F_k$ from the request input and output lengths; we obtain $Q$ as the \textit{normalized latency}—the end-to-end sequence latency divided by its output length, following prior work~\cite{kwon2023vllm,yu2022orca,jeong2025accelerating}. 

We fit the QoE‐model parameters $D_k$ by profiling. We begin by partitioning requests in the dataset into exponentially growing length buckets (e.g.\ $[100,200)$, $[200,400)$, $[400,800)$, $\dots$), discarding those that are too short. For each bucket, we measure online inference latency under batch sizes $1,2,4,\dots$ up to the maximum supported by memory, thereby controlling the request rate within a fixed memory budget. To profile a batch size $B$, we continuously inject requests to maintain $B$ in-flight; whenever one request completes, another is enqueued. Each $(\text{length},B)$ configuration runs for a fixed duration. From the resulting trace, we extract each request's normalized latency $Q$ and its average batch loads $F_k$ over its lifetime, then perform a least-squares regression of $Q$ against $F_k$ to estimate $D_k$. Denoting by $N$ the total number of requests, by $Q^{(j)}$ the normalized latency of the $j$th request, and by $F_k^{(j)}$ its $k$th batch-load feature, we compute
$$
\bigl(\hat D_0,\dots,\hat D_4\bigr)
=\arg\min_{D_0,\dots,D_4}
\sum_{j=1}^N\Bigl(Q^{(j)}-\sum_{k=0}^4 D_k\,F_k^{(j)}\Bigr)^2.
$$
For validation, we split the dataset into a fitting set and a validation set, and apply the same benchmarking procedure to the validation set as used during profiling. We compute the relative prediction error and show the results in \cref{ssec:eval-ablation}.

\subsection{Length-Aware Stage Partition}
\label{ssec:pipeline_planning}

\parab{Problem formulation.}
Finer-grained pipelines with more stages reduce length heterogeneity within each stage but incur higher inter-stage communication overhead. Conversely, allocating too many instances to a stage may reduce the number of stages and increase intra-instance sequence-length heterogeneity. A well-designed pipeline must balance these factors.

Pipeline planning runs at initialization and periodically thereafter. \sysname collects request statistics over the period (input and output lengths) and generates pipeline settings for the next period. 

The pipeline planning problem is to find optimal pipeline settings (number of stages $s$, number of instances per stage, and range of each stage) given $E$ instances and a set of requests. Let $n_{l',l}$ denote the subset of requests whose sequence lengths lie in $[l',l)$. The pipeline's ``goodness'' is quantified as the total QoE of all instances processing all requests, called pipeline quality.

\parab{Algorithm.} Let $f_{s,e,l}$ denote the optimal pipeline quality of serving all sequences with length $\le l$ using $s$ stages and $e$ instances. $f_{s,e,l}$ can be recursively represented as the sum of the optimal quality of the previous $s-1$ stages and the QoE of the $s$-th stage; enumerating all possible numbers of instances and requests in the $s$-th stage gives the optimal $f_{s,e,l}$:
\[
f_{s,e,l} = \min_{e' \in [s-1, e],\; l' \in [0, l-1]} f_{s-1,e',l'} + (e-e')Q^{n_{l',l}/(e-e')} + c_{l'}.
\]
$c_{l'}$ captures the delay of transferring all sequence fragments straddling the cut at length $l'$, computed based on the volume of transferred tokens and system bandwidth. 
$n_{l',l}/(e-e')$ means dividing the request set $n_{l',l}$ evenly among $e-e'$ subsets\footnote{We approximate a subset of the set division $S/n$ by sorting elements in the set, starting from the $n/2$-th one, and picking every $n$-th element to comprise the subset.}.
$Q^{n_{l',l}/(e-e')}$ is the QoE of one instance in its request processing, computed by \cref{eq:batch_quality}.

The recursion can be solved by dynamic programming. Finding the smallest value among all $f_{*,E,*}$ (all possible numbers of stages $s$ and sufficiently large sequence length $L$) gives the optimal pipeline settings.

\begin{figure}
    \centering
    \includegraphics[width=\linewidth]{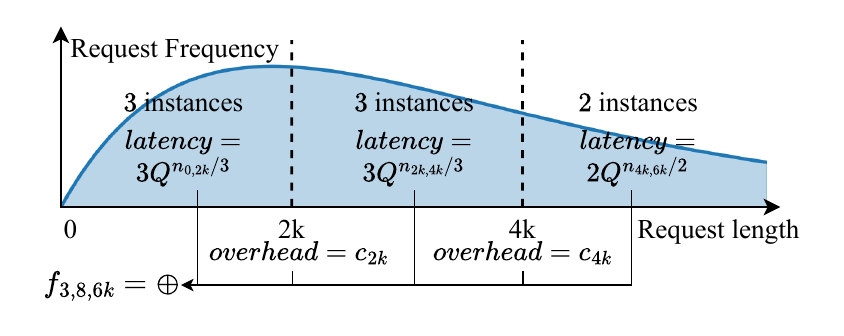}
    \caption{Pipeline planning based on the request length distribution. }
    \label{fig:topo-example}
\end{figure}

For example, \cref{fig:topo-example} illustrates the composition of $f_{3,8,6k}$ for a three-stage pipeline with $E=8$ instances. The DP algorithm selects cut points at lengths 2k and 4k, allocating 3, 3, and 2 instances to stages covering lengths $[0,2k)$, $[2k,4k)$, and $[4k,6k)$, respectively. The QoE of each stage is $3 Q^{n_{0,2k}/3}$, $3 Q^{n_{2k,4k}/3}$, and $2 Q^{n_{4k,6k}/2}$, and the inter-stage migration overheads at the cut points are $c_{2k}$ and $c_{4k}$. Summing these three QoEs and the two overheads yields the final estimate.

\parab{Optimizing efficiency.}
Naively evaluating all $(s,e,l)$ triples incurs a time complexity of $O(E^3L^2)$, even with $O(1)$ prefix-sum access to $n$ and $T$. To make this dynamic program tractable, we introduce two key optimizations. 
First, we bucket sequence lengths into exponentially increasing tiers (e.g., $[1,2)$, $[2,4)$, $[4,8)$, $\ldots$), reducing the number of candidate cut points from $L$ to $O(\log L)$. This is motivated by the  observation that real-world inference workloads are highly skewed: most sequences are short or moderate in length, while very long sequences are rare and sparsely distributed. Coarse-grained binning in the long-tail region prevents a few outliers from dominating the DP state space, reducing complexity from $O(E^3L^2)$ to $O(E^3 \log^2 L)$ while retaining modeling fidelity.

Second, we apply a two-phase heuristic. We first run a simplified DP that assigns exactly one instance per stage, yielding an initial $E$-stage pipeline in $O(E \log^2 L)$ time. We then iteratively merge adjacent stages to reduce total latency. For each pair, we define a \emph{merge gain}—the reduction in latency from unifying their instance and sequence range—and greedily merge the pair with the highest positive gain. Gains are tracked in a max-heap for $O(E \log E)$ updates, or computed via a naive $O(E^2)$ scan. This process continues until no further improvement is possible.

These optimizations reduce the end-to-end complexity to $O(E(\log^2 L + \log E))$, enabling efficient pipeline planning at cluster scale. While the number of instances $E$ per model is typically small (e.g., a few dozen), the supported context length $L$ can reach 128K tokens or more, making algorithmic efficiency crucial.

\subsection{Adaptive Range Refinement}
\label{ssec:boundary_adjustment}

\SetKwInOut{Input}{Input}

Each instance periodically refines its range boundary based on LoadTracker. At each refinement interval, the instance collects the sequence lengths from both itself and its successors. \sysname first computes the average workload of the next stage: it merges the successor sequence lengths as a union set, and divides the set evenly by the number of successors (using the same set division method in \cref{ssec:pipeline_planning}).

\sysname merges the instance's local sequence lengths and the average successor lengths into a single set. It sorts the requests in the set as a list $R$ with requests indexed from $0$ to $N-1$. \sysname enumerates all possible splitting points in $R$, computes the overall quality of both parts by \cref{eq:batch_quality}, and finds the optimal one with the best overall quality:
$$
b=\arg\min_{0\le i< N}\;\left( Q^{R[:i]}+Q^{R[i:]}\right).
$$
Thus, the boundary is identified as the length at the optimal splitting point $R[b]$.

To ensure boundary stability and prevent performance degradation caused by oscillating partitions, we apply three optimizations. First, the boundary is initialized using the offline pipeline planning result, providing a high-quality starting point that accelerates convergence. Second, we smooth the boundary updates using exponential moving averaging. This acts as a low-pass filter to dampen sensitivity to short-term workload fluctuations, thereby avoiding spurious transfers and unnecessary adjustments driven by transient spikes. Third, we suppress refinements during periods of low traffic (e.g., when there are fewer than five requests). In such sparse regimes, the arrival or completion of a single request can disproportionately skew the length distribution; freezing the boundary prevents these discrete events from causing huge shifts in partition logic, ensuring reliable decisions.

\subsection{Decentralized Load (Re)Balancing}
\label{ssec:bid_ask}

\begin{figure}[t]
    \centering
    \includegraphics[width=\linewidth]{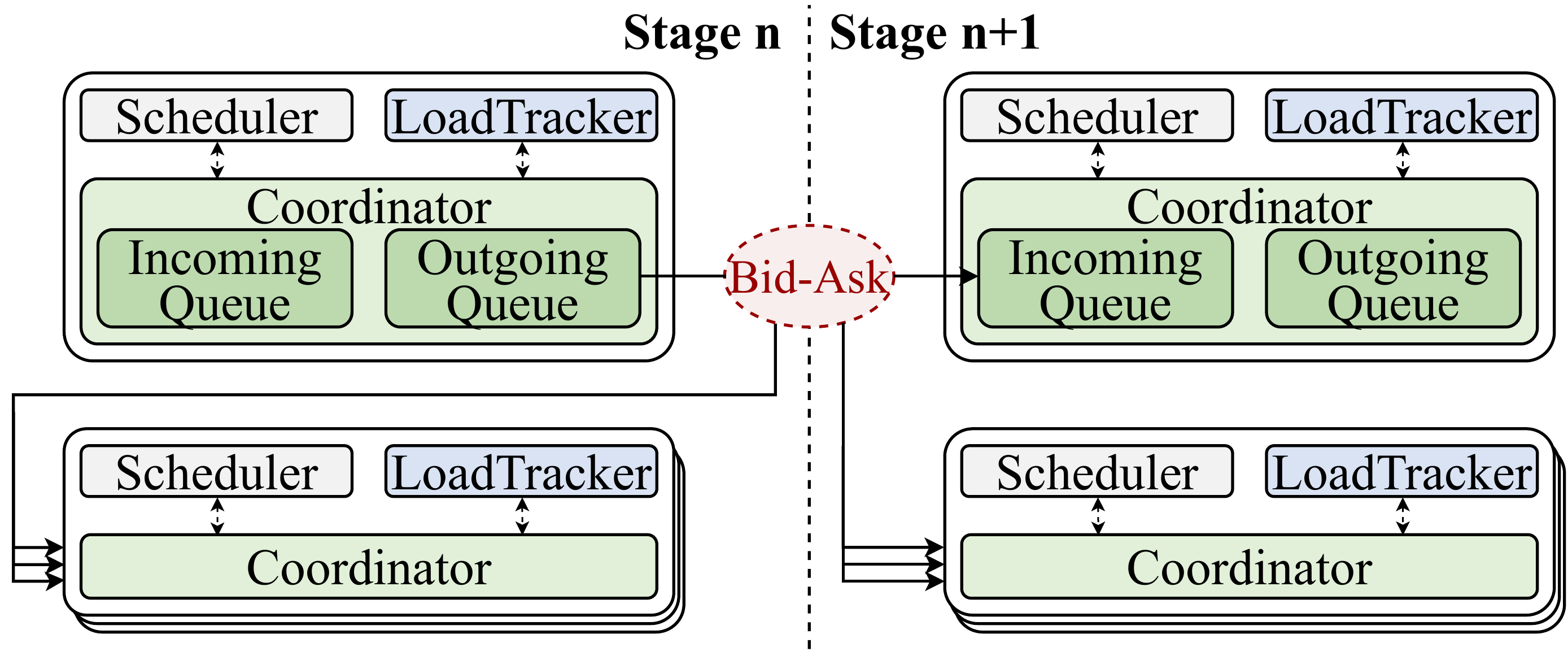}
    \caption{Illustration of intra‑stage load balancing using dynamic decentralized bid-ask scheduling.}
    \label{fig:bidask}
\end{figure}

\parab{Two classes of intra-stage load (re)balancing.} When an upstream instance hands over requests to its downstream successors, the request distribution should be balanced. This inter-stage request handover triggers bid-ask scheduling.

When instances within one stage execute for a while, their loads may diverge due to diverse requests, requiring rebalancing. With instance LoadTracker exchanging their workloads (sequence lengths), once an instance finds itself an overloaded outlier (e.g., its request-memory demand is 25\% above the stage average), it triggers intra-stage bid-ask scheduling. Both classes share the same protocol and migration subsystem below.

\parab{Bid-ask protocol.} For scalability, \sysname employs a decentralized bid-ask scheduling protocol: each sender distributes its requests to receivers independently, and all senders' requests are balanced across all receivers.

Senders and receivers negotiate pairwise for request handover, akin to transaction matching in financial markets. This analogy inspires a lightweight \emph{bid-ask protocol}~\cite{glosten1985bid} to match senders and receivers without global coordination.

The protocol works as follows. In the ``asking'' phase, the sender notifies all candidate receivers of one request migration. In the ``bidding'' phase, each receiver replies with its current load and the earliest time it can start the request migration (defined below). The sender filters out half the receivers with higher load, then keeps the three receivers with the earliest transmission start times, and selects the one that replies first. The sender notifies the chosen receiver to confirm the ownership handover, including request metadata such as request ID, sequence length, and source instance. The sender continues to hand over other requests.

\parab{Request migration.}
\cref{fig:bidask} illustrates the transmission subsystem.
The sender buffers all requests to migrate.
The receiver buffers the requests it wins in a \textit{priority queue}.

In the bid-ask protocol, the sender's asking message piggybacks its load (the total length of all buffered requests), and the receiver's bidding message piggybacks its earliest transmission start time (the total length of its buffered requests divided by its measured throughput).
A request's priority is defined as the sender's load, which is used to order the requests in the priority queue.

The receiver repeatedly dequeues the highest-priority request from the priority queue and initiates migration. If the sender is transmitting another request, the receiver skips this request and attempts the next; otherwise, it starts transmission immediately.

A receiver's first request in the priority queue could starve if the sender repeatedly transmits other requests. To avoid this, the receiver records the number of failed attempts for each request; once a request fails more times than a threshold, the receiver notifies the sender about the starvation, prompting the sender to send the request immediately after completing the current transmission. The receiver then waits to receive the request instead of attempting the next one.

For request transmission, we adopt a \emph{live-migration} mechanism proposed by Llumnix~\cite{sun2024llumnix}, which pipelines KV-cache transfer with ongoing decoding in multiple rounds, minimizing disruption to the source instance. When a transmission completes, the sender removes the request from its buffer and deletes the sequence, and the receiver removes the request from its priority queue and hands the sequence to the local scheduler for further processing.

\begin{figure*}[t]
    \centering
    \includegraphics[width=\linewidth]{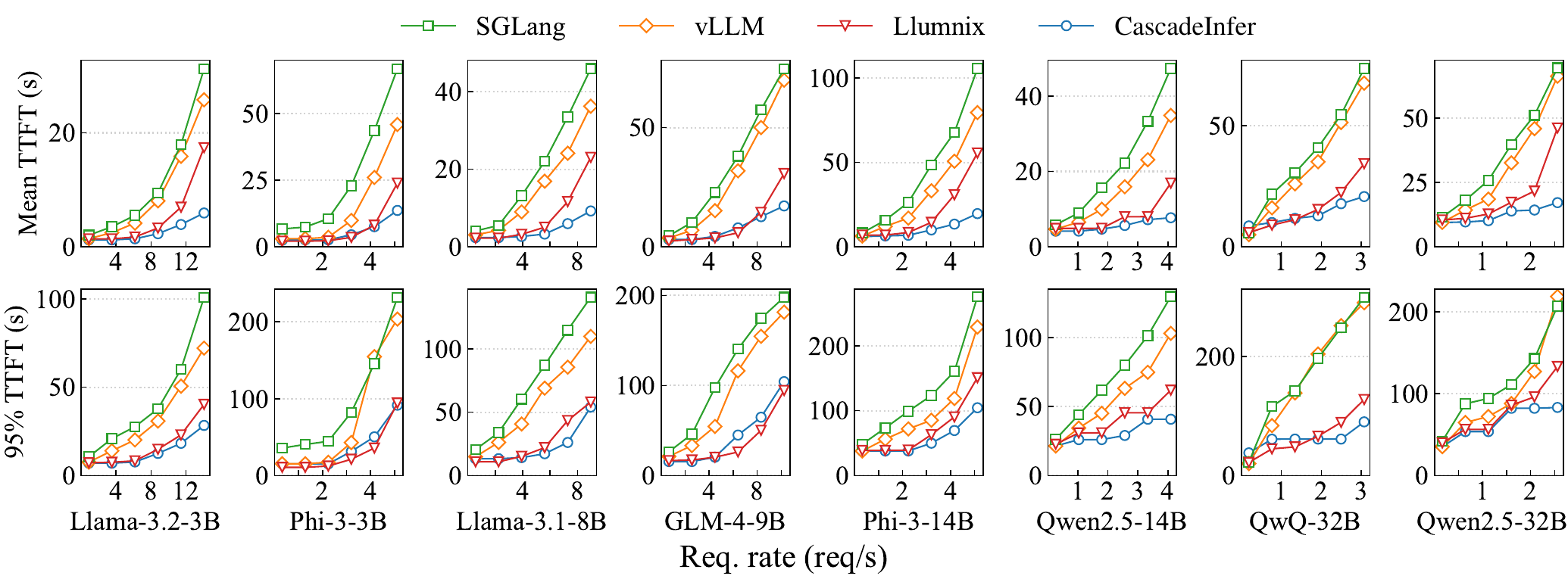}
    \caption{Mean and 95th-percentile TTFT measured across different LLM models under varying request arrival rates.}
    \label{fig:ttft}
\end{figure*}

\section{Implementation}
\label{sec:impl}
\sysname is implemented with ~3,000 lines of Python for the control plane and 700 lines of C++ for the high-performance communication backend, built atop vLLM (v0.9.1) without modifying its core attention kernels. We integrate \sysname modularly into vLLM via Scheduler-Aware Hooks: the LoadTracker monitors token-level workloads, while the Coordinator manages metadata and orchestrates KV cache migrations. 

\noindent \textbf{Efficient migration infrastructure.} To minimize the overhead of frequent, fine-grained KV transfers, we implement a custom C++ communication backend that bypasses Python’s GIL and avoids the latency of NCCL’s collective operations, which are ill-suited for numerous small messages. Instead, the backend leverages low-level APIs—specifically \textit{cudaMemcpyPeerAsync} for intra-node and RDMA verbs for inter-node transfers—to achieve zero-copy GPU-to-GPU transmission. Furthermore, \sysname optimizes physical topology by placing instances of adjacent pipeline stages on the same node whenever possible, thereby significantly reducing expensive inter-node communication.

\noindent \textbf{Migration flow control and optimization.} \sysname optimizes runtime execution by eliminating intermediate buffers; KV caches are transferred directly into idle slots on the target instance, and migration is skipped if no idle cache is available. To prevent bandwidth contention, \sysname enforces a strict concurrency limit (capped at three parallel transfers in our implementation); requests exceeding this threshold continue running on the source to avoid performance regression. Finally, we employ asynchronous multi-round live migration (adapting Llumnix~\cite{sun2024llumnix}) combined with bidirectional transfer support and a starvation-aware backpressure mechanism. This design effectively overlaps communication with computation, preventing head-of-line blocking while maximizing GPU and NIC utilization.

\section{Evaluation}
\label{sec:evaluation}

This section addresses four key questions:

\noindent
(1) How does \sysname affect request latency across diverse configurations? (\cref{ssec:eval-latency})

\noindent
(2) How does \sysname influence system throughput across diverse configurations? (\cref{ssec:eval-throughput})

\noindent
(3) How does \sysname impact SLO attainment? (\cref{ssec:eval-slo})

\noindent
(4) What is the effect of each \sysname mechanism? (\cref{ssec:eval-ablation})

\subsection{Experimental Setup}
\label{ssec:exp_setup}

\parab{Testbed.} We evaluate \sysname on two testbeds, each comprising two nodes with 16 NVIDIA GPUs. The first testbed uses NVLink-connected H20 GPUs, each with 141~GB of memory, while the second employs PCIe-connected L40 GPUs, each with 48~GB. Every node in both setups is equipped with 2~TB of host memory, an Intel\textregistered~Xeon\textregistered~Platinum 8468 CPU (192 logical cores), and a 400~Gbps ConnectX-7 NIC.

\parab{Baselines.} We compare \sysname against three popular open-source inference frameworks: vLLM 0.9.1~\cite{kwon2023vllm}, SGLang 0.4.9~\cite{sglang}, and Llumnix~\cite{sun2024llumnix}. vLLM and SGLang operate as standalone inference service instances without inter-instance scheduling; we therefore deploy a simple round-robin load balancer across their instances. In contrast, Llumnix extends vLLM with an integrated inter-instance scheduler, which we run using its default configuration. Unless stated otherwise, all backends use an FCFS policy. We cap the batch size at 1024—the vLLM official default—with actual batch sizes further constrained by memory and token budgets at runtime. Following the official default settings, we use FlashAttention~\cite{dao2022flashattention} 
backend for vLLM, Llumnix, and \sysname, and use FlashInfer~\cite{ye2025flashinfer} backend for SGLang~\cite{sglang}.

\parab{Models.} We evaluate eight LLMs in FP16 precision across a range of commonly used sizes, grouped into four categories: (1) \textit{Tiny} — Llama-3.2-3B~\cite{llama-3.2} and Phi-3-3B~\cite{team2024phi}; (2) \textit{Small} — Llama-3.1-8B~\cite{llama-3.1} and GLM-4-9B~\cite{glm2024chatglm}; (3) \textit{Moderate} — Phi-3-14B~\cite{team2024phi} and Qwen-2.5-14B~\cite{qwen2025qwen25technicalreport}; and (4) \textit{Large} — QwQ-32B~\cite{qwq} and Qwen-2.5-32B~\cite{qwen2025qwen25technicalreport}. Each instance runs on a single GPU, except for Llama-3.1-70B~\cite{llama-3.1}, which we evaluate in 2- and 4-way tensor-parallel configurations.\footnote{
We make homogeneous instance setups and leave the discussion of heterogeneous configuration in \cref{sec:discussion}.
} 
We exclude larger models because their high per-instance GPU requirements would leave too few instances on our testbeds, misrepresenting typical multi-instance deployments. \sysname constructs pipelines of 4 to 6 stages for the selected model, with each stage comprising 1 to 4 instances. All selected models support context lengths of at least 128K tokens.

\parab{Workload.} Following the standard practice in prior work~\cite{kwon2023vllm,wu2023fast,sun2024llumnix}, we construct workloads using the ShareGPT~\cite{shareGPT} datasets of ChatGPT conversation histories and model request arrivals as a Poisson process with configurable rates. Each test point runs for the same duration.

\begin{figure*}[t]
    \centering
    \includegraphics[width=\linewidth]{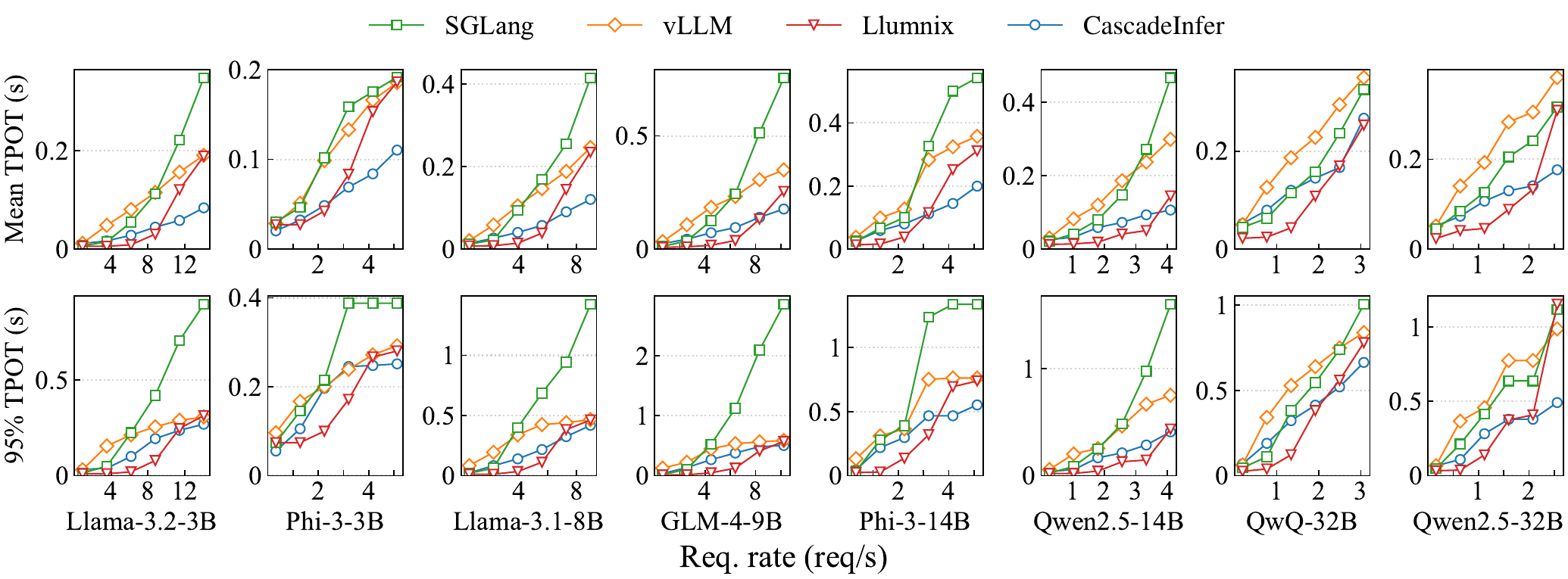}
    \caption{Mean and 95th-percentile TPOT measured across different LLM models and varying request arrival rates.}
    \label{fig:tpot}
\end{figure*}

\parab{Metrics.} We evaluate inference performance using four key metrics: \textit{Time to First Token (TTFT)} — the time from request submission to the generation of the first output token; \textit{Time per Output Token (TPOT)} — the average latency to generate each subsequent token in a request; \textit{Throughput} — system-wide token generation rate; and \textit{SLO Attainment} — the percentage of requests meeting the defined SLO.

\parab{Experiment parameters.} We vary request arrival rates to cover both light and heavy loads. Light load verifies that \sysname does not introduce a negative impact, while heavy load, our target scenario, evaluates its performance gains.

\subsection{Reduced End-to-End Latency}
\label{ssec:eval-latency}

\parab{Latency reduction across models.} We compare \sysname with other systems across eight LLMs on the H20 testbed. The results are shown in Figures~\ref{fig:ttft} and \ref{fig:tpot}.

\begin{figure}[t]
    \includegraphics[width=\linewidth]{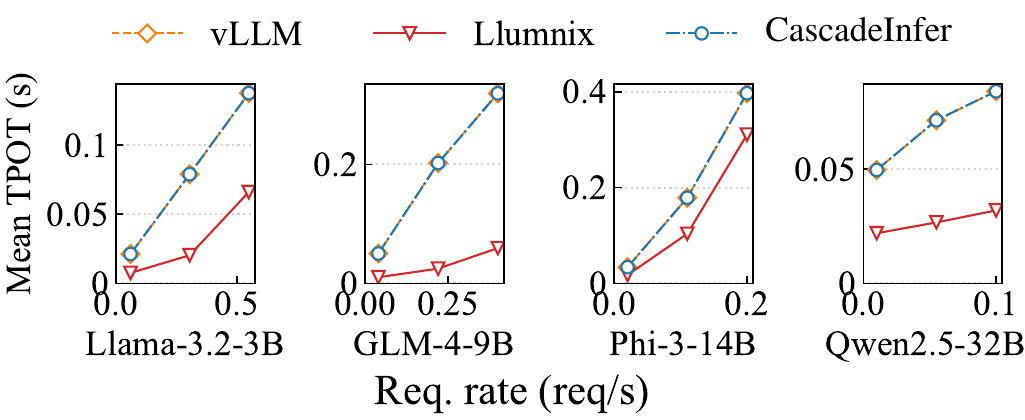}
    \caption{TPOT of a \textit{single} instance across varying request arrival rates. \sysname's single-instance performance matches vLLM's but falls behind Llumnix's. By comparing this to other results in \S\ref{sec:evaluation}, we find that \sysname's multi-instance scheduling delivers higher gains than Llumnix's.}
    \label{fig:single-instance}
\end{figure}

\noindent
\underline{\emph{TTFT.}} 
As shown in \cref{fig:ttft}, \sysname consistently achieves a lower mean TTFT than vLLM, SGLang, and Llumnix. Under heavy load, it reduces mean TTFT by 67\%–78\% compared to vLLM, by 70\%–84\% compared to SGLang, and by 36\%–66\% compared to Llumnix, the state-of-the-art in multi-instance scheduling system. 
Tail TTFT exhibits similar trends, with average reductions of 56\%, 62\%, and 19\%, respectively. 
Under light load, model access overhead (unrelated to requests) dominates latency, so the system reverts to standard multi-instance scheduling with performance comparable to Llumnix.

\noindent
\underline{\emph{TPOT.}} 
As shown in \cref{fig:tpot}, \sysname achieves a lower mean TPOT than vLLM, SGLang, and Llumnix in most settings. Under heavy load, it reduces mean TPOT by 30\%–64\% vs.\ vLLM, 25\%–77\% vs.\ SGLang, and 3.4\%–56\% vs.\ Llumnix. 
\sysname also reduces tail TPOT by 25\% vs.\ vLLM, 61\% vs.\ SGLang, and 20\% vs.\ Llumnix under heavy load.

Under light load, Llumnix performs better because it runs on a newer engine with lower latency. Since \sysname focuses on scheduling requests across engine instances rather than optimizing a single engine instance, its benefits here are actually underestimated—on that same newer engine, \sysname would achieve even lower TPOT. We confirmed this by experiments below. The results show that \sysname's single-instance performance is on par with vLLM's but falls short of Llumnix's.

\begin{figure}[t]
    \centering
    \includegraphics[width=1.0\linewidth]{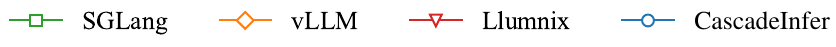}
    \\
    \begin{subfigure}[t]{0.47\linewidth}
        \includegraphics[width=\linewidth]{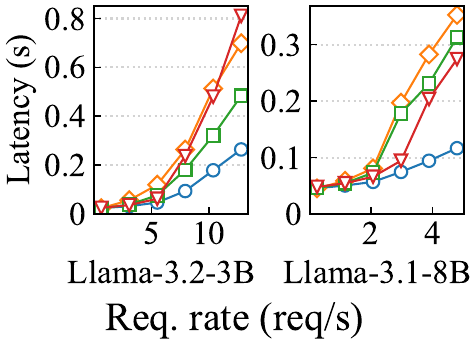}
        \caption{L40 testbed}
        \label{fig:l40-latency}
    \end{subfigure}
    \hfill
    \begin{subfigure}[t]{0.47\linewidth}
        \includegraphics[width=\linewidth]{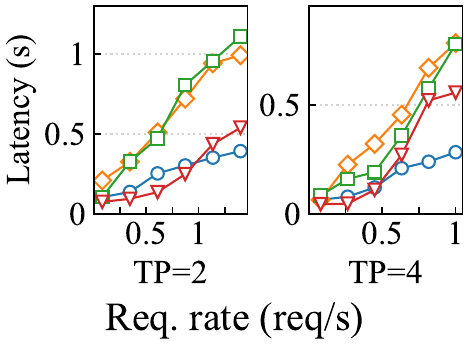}
        \caption{Tensor parallelism}
        \label{fig:tp-latency}
    \end{subfigure}
    \caption{Normalized latency measured on an additional testbed and tensor-parallelism configurations. }
    \label{fig:extra-latency-}
\end{figure}

\begin{figure*}[t]
    \centering
    \includegraphics[width=\linewidth]{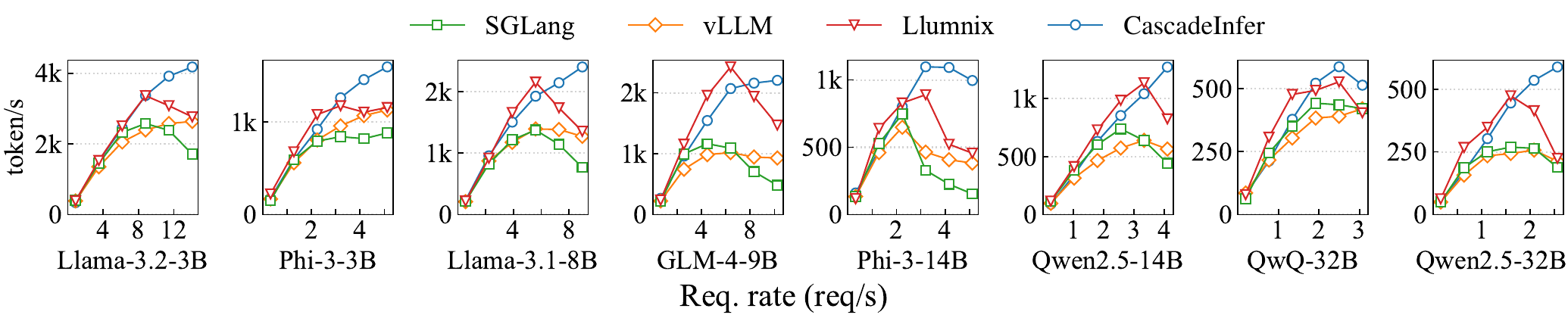}
    \caption{System throughput measured across different LLM models under varying request arrival rates.}
    \label{fig:throughput}
\end{figure*}

\textit{Comparison of single-instance performance.}
\sysname focuses on improving cluster‐level performance by using length‐aware request scheduling across instances. When processing requests from the same batch, \sysname delivers single‐instance performance on par with vLLM, demonstrating full compatibility with both existing and future intra‐instance optimizations. In a side-by-side comparison of single-instance performance among \sysname, vLLM, and Llumnix (\cref{fig:single-instance}), \sysname matches vLLM but is still outperformed by Llumnix's latest engine by 22\%--81\% since Llumnix's engine has a lower runtime scheduling and executing overhead than vLLM. This gap indicates that the cluster‐level gains reported in the evaluation actually underestimate \sysname's potential—when paired with the newest engine, \sysname can achieve even higher overall performance.

\parab{Latency reduction across testbeds.} To demonstrate that \sysname's performance gains are platform-independent, we evaluate it on the L40 testbed, which has 48~GB of memory and thus only supports small models. We measure normalized latency, defined as the end-to-end delay per output token, including both TTFT and TPOT. As shown in \cref{fig:l40-latency}, \sysname outperforms all baselines and reduces latency by 45\%–67\%.

\parab{Latency reduction under tensor parallelism.} To evaluate \sysname on large models spanning multiple GPUs, we run Llama-3.1-70B with tensor parallelism (TP) of 2 and 4 on the H20 testbed. At TP=2, model parameters occupy nearly half of each GPU's memory, causing higher parameter access latency and reducing \sysname's effectiveness compared to TP=4. As shown in \cref{fig:tp-latency}, \sysname reduces latency by 27\%–65\% with 2 GPUs and 49\%–64\% with 4 GPUs, outperforming all baselines.

\begin{figure}[t]
    \centering
    \includegraphics[width=1.0\linewidth]{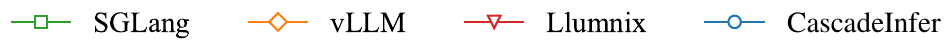}
    \\
    \begin{subfigure}[t]{0.47\linewidth}
        \includegraphics[width=\linewidth]{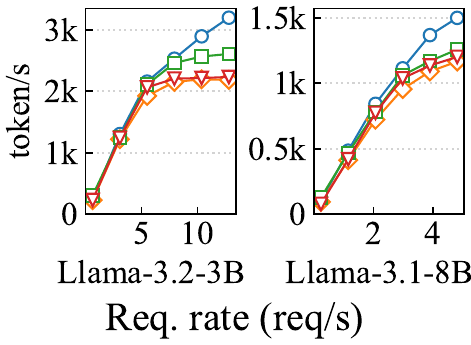}
        \caption{L40 testbed}
        \label{fig:l40-throughput}
    \end{subfigure}
    \hfill
    \begin{subfigure}[t]{0.47\linewidth}
        \includegraphics[width=\linewidth]{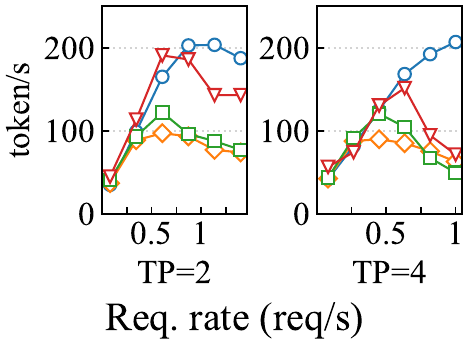}
        \caption{Tensor parallelism}
        \label{fig:tp-throughput}
    \end{subfigure}
     \caption{System throughput measured on an additional testbed and tensor-parallelism configurations.}
    \label{fig:extra-throughput}
\end{figure}

\subsection{System Throughput Improvement}
\label{ssec:eval-throughput}

\parab{Throughput improvement across models.} \cref{fig:throughput} shows the throughput of each system across eight models on the H20 testbed. Throughput declines once the request rate exceeds a threshold, as all systems prioritize prefilling over decoding. \sysname sustains a higher threshold than all baselines. Under heavy load, its average throughput reaches 1.99$\times$ and 2.18$\times$ those of vLLM and SGLang, respectively, and is 1.71$\times$ that of Llumnix. These gains stem from \sysname's improved attention backend efficiency by reducing length heterogeneity.

\parab{Throughput improvement across testbeds.} \cref{fig:l40-throughput} shows system throughput on the L40 GPU. Under heavy load, \sysname achieves average throughputs of 1.37$\times$, 1.21$\times$, and 1.34$\times$ those of vLLM, SGLang, and Llumnix, respectively. These gains are smaller than on H20, as the L40's lower memory caps batch sizes and narrows the length-heterogeneity gap.

\parab{Throughput improvement under tensor parallelism.} As shown in \cref{fig:tp-throughput}, \sysname achieves 1.31–2.53$\times$ throughput with 2 GPUs and 2.89–4.16$\times$ throughput with 4 GPUs compared to the baselines. The benefit increases with TP size, since tensor parallelism reduces each GPU's model-access overhead.

\begin{figure}[t]
    \centering
    \includegraphics[width=1.0\linewidth]{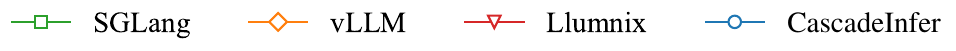}
    \\
    \begin{subfigure}[t]{0.32\linewidth}
        \includegraphics[width=\linewidth]{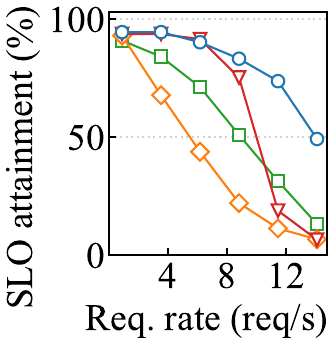}
        \caption{5$\times$ SLO}
        \label{fig:slo-1}
    \end{subfigure}
    \hfill
    \begin{subfigure}[t]{0.32\linewidth}
        \includegraphics[width=\linewidth]{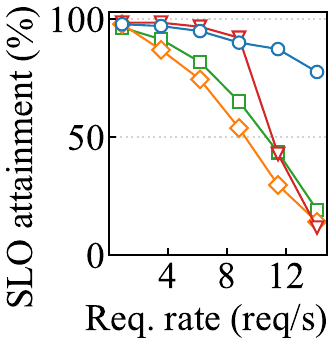}
        \caption{10$\times$ SLO}
        \label{fig:slo-2}
    \end{subfigure}
    \hfill
    \begin{subfigure}[t]{0.32\linewidth}
        \includegraphics[width=\linewidth]{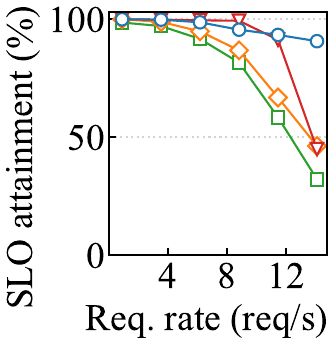}
        \caption{20$\times$ SLO}
        \label{fig:slo-3}
    \end{subfigure}
    \caption{SLO attainment measured across SLO levels and varying request arrival rates. The SLO is defined by the mean TTFT and TPOT at minimum system load, and the N$\times$ SLO scales both constraints N times. } 
    \label{fig:slo}
\end{figure}

\subsection{Improved Service Level Objective Attainment}
\label{ssec:eval-slo}

Service Level Objectives (SLOs) define worst‐case bounds on key inference‐service metrics (e.g., TTFT, TPOT) to ensure a consistent user experience. We quantify SLO attainment as the fraction of those bounds actually met—higher attainment indicates better service quality. We define the baseline SLO as the TTFT and TPOT under minimum load (one request), then evaluate attainment when both bounds are scaled by a factor of N. \cref{fig:slo} shows the results of Llama-3.2-3B on the H20 testbed; results for other models, omitted for brevity, exhibit similar trends. Compared with the baselines, \sysname achieves 3.8–7.6$\times$ attainment under 5$\times$ SLO and 2.0–2.8$\times$ attainment under 20$\times$, driven by its lower TTFT and TPOT.

\begin{figure}[tb]
    \centering
    \includegraphics[width=0.51\linewidth, valign=t]{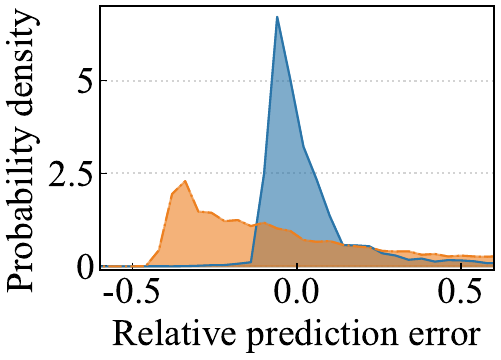}
    \includegraphics[width=0.45\linewidth, valign=t]{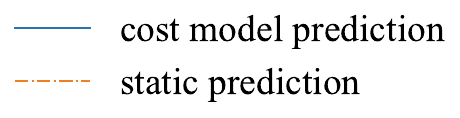}
    \caption{Prediction error of our cost model. Errors closer to zero are better.} 
    \label{fig:prediction}
\end{figure}

\begin{figure}[t]
    \centering
    \includegraphics[width=0.72\linewidth]{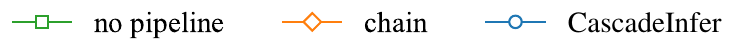}
    \\
    \begin{subfigure}[t]{0.49\linewidth}
        \centering
        \includegraphics[width=0.65\linewidth]{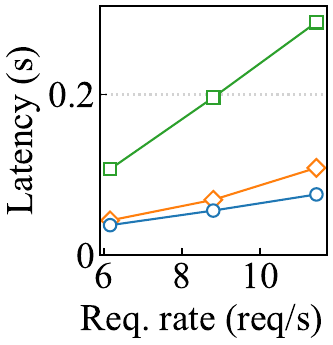}
        \caption{Normalized latency}
        \label{fig:topo-latency}
    \end{subfigure}
    \begin{subfigure}[t]{0.49\linewidth}
        \centering
        \includegraphics[width=0.67\linewidth]{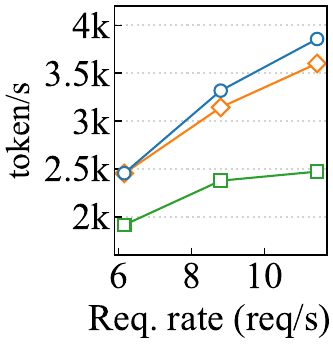}
        \caption{System throughput}
        \label{fig:topo-throughput}
    \end{subfigure}
    \caption{Performance across layouts and varying request arrival rates. }
    \label{fig:topo}
\end{figure}

\subsection{Design Choice Validation and Ablation Study}
\label{ssec:eval-ablation}

Our ablation study yielded consistent observations and conclusions across all models and testbeds; for simplicity, we present results for Llama-3.2-3B on the H20 testbed.

\parab{Accuracy of the QoE model for latency estimation.} As the basis for pipeline planning and boundary adjustment, we first validate the accuracy of our QoE model's latency estimates. We perform profiling experiments across various request lengths and rates to calibrate the model parameters, and then conduct a validation study using these parameters. As a baseline, we use a static predictor that always returns the global mean latency. \cref{fig:prediction} shows the probability density of per-request relative prediction errors. Our model's error distribution peaks sharply at zero, with an average error of 8.9\%, versus 64\% for the static baseline.

\parab{Length-aware vs. na\"ive stage partitioning.} 
We compare \sysname with two na\"ive layouts: a \textit{chain} layout, which assigns one instance per pipeline stage, and a \textit{no-pipeline} layout, which coalesces all instances into a single stage. As shown in \cref{fig:topo}, the absence of length-based partitioning causes the no-pipeline design to perform the worst, while the chain layout underperforms \sysname due to its overly long pipeline, which incurs higher migration overhead and complicates load balancing across stages. Consequently, \sysname achieves 30\% lower latency and 7.1\% higher throughput compared to the chain layout.

\begin{figure}[t]
    \centering
    \includegraphics[width=0.95\linewidth]{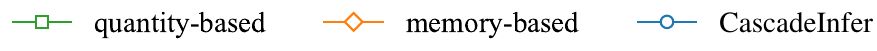}
    \\
    \begin{subfigure}[t]{0.49\linewidth}
        \centering
        \includegraphics[width=0.65\linewidth]{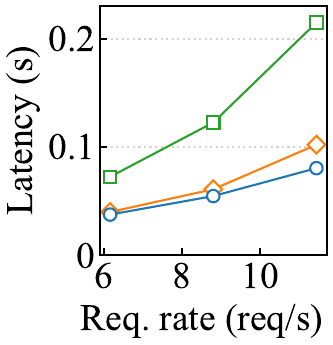}
        \caption{Normalized latency}
        \label{fig:partition-latency}
    \end{subfigure}
    \begin{subfigure}[t]{0.49\linewidth}
        \centering
        \includegraphics[width=0.67\linewidth]{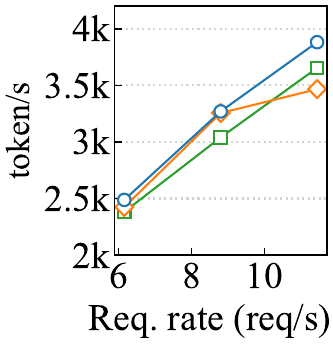}
        \caption{System throughput}
        \label{fig:partition-throughput}
    \end{subfigure}
    \caption{Performance across boundary-adjustment policies and varying request arrival rates.}
    \label{fig:partition}
\end{figure}

\parab{Adaptive vs. na\"ive range-refinement policies.} 
We compare \sysname against two na\"ive range-refinement policies: a quantity-based refinement that balances the number of requests per stage, and a memory-based refinement that equalizes per-stage memory usage. Both policies enforce increasing request lengths across stages, as in \sysname. However, the quantity-based approach yields severe workload imbalance and thus the poorest performance, while the memory-based scheme fails to account for request heterogeneity as in \cref{ssec:boundary_adjustment}. As \cref{fig:partition} illustrates, \sysname outperforms the memory-based policy by 21\% lower latency and 12\% higher throughput.

\parab{Complexity reduction for stage partitioning.} With 16 instances and a 128K sequence length, our optimized algorithm (\cref{ssec:pipeline_planning}) completes stage partitioning in 0.06s, versus an estimated 51 hours without optimization, yielding a speedup of roughly $3\times 10^6$.

\begin{figure}[tb]
    \centering
    \includegraphics[width=0.51\linewidth, valign=t]{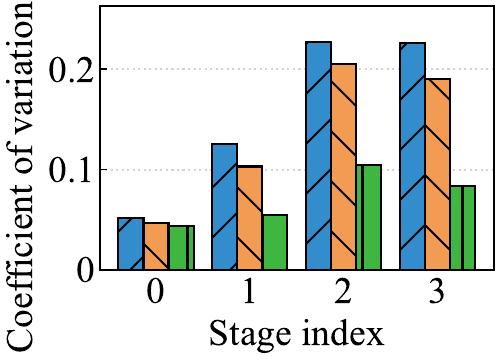}
    \includegraphics[width=0.4\linewidth, valign=t]{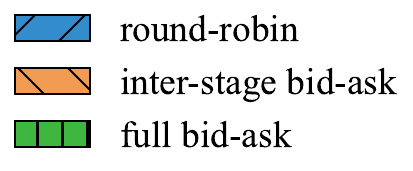}
    \caption{Imbalance in output token counts across instances at each stage. Lower values are better.}
    \label{fig:load-balancing}
\end{figure}

\parab{The load-balancing effect of bid-ask protocol.} We construct a four-stage pipeline with four instances per stage and compare three load-balancing policies. For each stage, we record the number of output tokens per instance and compute the coefficient of variation (CV), with lower values indicating a more balanced load. As shown in \cref{fig:load-balancing}, the round-robin policy exhibits the highest imbalance, since it dispatches requests regardless of instance load. Inter-stage-only bid-ask scheduling reduces this imbalance, while full bid-ask scheduling further mitigates the variability from unpredictable output lengths, yielding an average CV reduction of 40\% over inter-stage-only bid-ask and 47\% over round-robin.

\section{Related Work}
\noindent\textbf{Attention backends.} Modern self-attention kernels such as FlashAttention~\cite{dao2022flashattention} and FlashInfer~\cite{ye2025flashinfer} improve memory efficiency via I/O-aware designs and paged KV caches~\cite{kwon2023vllm}. Transformer libraries~\cite{fasttransformer,xFormers2022} and low-level frameworks like Triton~\cite{tillet2019triton}, CUTLASS~\cite{cutlass}, and CUDA~\cite{guide2020cuda} provide flexible platforms for scalable attention. \sysname complements these by reducing request-level length heterogeneity to unlock the efficiency of the attention backend.

\noindent\textbf{Intra-instance scheduling.} vLLM~\cite{kwon2023vllm} and SGLang~\cite{sglang} dynamically form iteration batches using FCFS or length-aware policies. Prior work~\cite{zheng2023response} reduced output-length heterogeneity in traditional one-shot batching through output-length prediction, thereby reducing hardware idle time. However, this batching paradigm has been superseded by the more efficient continuous batching~\cite{yu2022orca} that underpins \sysname. FastServe~\cite{wu2023fast}, SSJF~\cite{qiu2024efficient}, and similar works~\cite{fu2024efficient} prioritize short-output requests to improve latency. Yet, this comes at the cost of throughput: effective prioritization often relies on small batch sizes (e.g., $\leq 10$ in SSJF) and aggressive preemption, leading to high swapping overhead~\cite{wu2023fast} that hinders system efficiency.
\sysname avoids this trade-off by distributing length-homogeneous batches across its instances. 

\noindent\textbf{Inter-instance scheduling and distributed inference.} Multi-instance systems~\cite{sun2024llumnix,dynamo,team2025aibrix} like Llumnix~\cite{sun2024llumnix} employ load- and cache-aware heuristics to balance traffic among instances. Others~\cite{zhong2024distserve,zhu2025megascale,step3} decouple inference into phases (e.g., prefill vs.\ decode) and assign these phases to specialized instance pools or heterogeneous hardware. \sysname is compatible with these approaches and enhances them via length-based instance partitioning.

\noindent\textbf{Request migration in LLM serving.} ServerlessLLM~\cite{fu2024serverlessllm} and Llumnix~\cite{sun2024llumnix} enable live request migration for stall-free handoff. \sysname builds on this by integrating migration into the decentralized bid-ask mechanism for balancing length-specialized instances.

\section{Discussion}
\label{sec:discussion}

This work exposes a fundamental limitation of current LLM serving: intra-instance schedulers operate under the constraints of Little's Law and cannot accumulate enough requests to mitigate sequence-length heterogeneity, while the attention backend is highly sensitive to such heterogeneity. This mismatch leads to persistent GPU underutilization even with optimized single-instance engines.

\noindent \textbf{Scope and limitations.} \sysname is most effective when ample GPU memory enables large-batch processing, shifting the primary latency bottleneck from parameter access to attention computation and KV cache intensity. Gains scale with available memory and parallelism. When the parameter size is large relative to the total GPU memory (e.g., occupying 80\% of the available capacity) and the remaining KV cache memory cannot support large batches, the realized benefits are naturally constrained, yet the method remains effective. Workloads with highly uniform lengths inherently present less heterogeneity, where \sysname still maintains correct scheduling and avoids overhead.

\noindent \textbf{Orthogonality.} \sysname is independent of any particular inference engine. While our prototype inherits vLLM’s single-instance performance characteristics, the scheduling layer itself is complementary to kernel-level and intra-instance optimizations; faster execution engines would only amplify the benefits of global coordination.

\noindent \textbf{Scalability.} \sysname combines offline bucket planning with lightweight online bid-ask rescheduling, keeping planning overhead low and KV migration infrequent. KV migration is efficient (\cref{sec:impl}) and rarely impacts performance under realistic bandwidth; even under stress, \sysname degrades gracefully.

\noindent \textbf{Heterogeneous instances.} While our evaluation employs homogeneous instance setups to align with standard production practices for simplified management, \sysname's mechanisms—including pipeline planning, boundary adjustment, and bid-ask scheduling—natively support heterogeneous configurations. However, fully exploiting this flexibility presents a significant challenge: modern inference engines (e.g., vLLM and SGLang) expose hundreds of configurable parameters (e.g., tensor, pipeline, and expert parallelism degrees). Identifying the optimal configuration for each pipeline stage involves a high-dimensional search with prohibitive profiling costs, which we leave for future exploration.
\section{Conclusion}
By elevating length heterogeneity from a local batching issue to a cluster-level scheduling problem, \sysname enables effective request regrouping across identical replicas, overcoming fundamental limits of per-instance schedulers. Through dynamic boundaries and decentralized rescheduling, \sysname achieves substantial utilization and throughput gains, advancing scalable and efficient LLM serving.

\bibliographystyle{plain}
\bibliography{ref}

\end{document}